\documentclass{emulateapj}

\shorttitle{Galaxy Environments with Deep Redshift Surveys}
\shortauthors{Cooper et al.}

\begin{document}



\title{Measuring Galaxy Environments with Deep Redshift Surveys}


\author{
Michael C.\ Cooper\altaffilmark{1},
Jeffrey A.\ Newman\altaffilmark{2,4}, 
Darren S.\ Madgwick\altaffilmark{2,4},
Brian F.\ Gerke\altaffilmark{3}, 
Renbin Yan\altaffilmark{1}, 
Marc Davis\altaffilmark{1,3}
}

\altaffiltext{1}{
Department of Astronomy, University of California at Berkeley, Mail Code
3411, Berkeley, CA 94720 USA; cooper@astro.berkeley.edu,
renbin@astro.berkeley.edu, marc@astro.berkeley.edu}  

\altaffiltext{2}{
Lawrence Berkeley National Laboratory, 1 Cyclotron Road Mail Stop 50-208,
Berkeley, CA 94720 USA; janewman@lbl.gov, dsmadgwick@lbl.gov}

\altaffiltext{3}{
Department of Physics, University of California at Berkeley, Mail Code
7300, Berkeley, CA 94720 USA; bgerke@astro.berkeley.edu}

\altaffiltext{4}{Hubble Fellow}

\begin{abstract}
We study the applicability of several galaxy environment measures
($n^{\rm th}$-nearest-neighbor distance, counts in an aperture, and
Voronoi volume) within deep redshift surveys. Mock galaxy catalogs are
employed to mimic representative photometric and spectroscopic surveys
at high redshift $(z \sim 1)$. We investigate the effects of survey
edges, redshift precision, redshift-space distortions, and target
selection upon each environment measure. We find that even optimistic
photometric redshift errors $(\sigma_z = 0.02)$ smear out the
line-of-sight galaxy distribution irretrievably on small scales; this
significantly limits the application of photometric redshift surveys
to environment studies. Edges and holes in a survey field dramatically
affect the estimation of environment, with the impact of edge effects
depending upon the adopted environment measure. These edge effects
considerably limit the usefulness of smaller survey fields (e.g.\ the
GOODS fields) for studies of galaxy environment. In even the poorest
groups and clusters, redshift-space distortions limit the
effectiveness of each environment statistic; measuring density in
projection (e.g.\ using counts in a cylindrical aperture or a
projected $n^{\rm th}$-nearest-neighbor distance measure)
significantly improves the accuracy of measures in such over-dense
environments. For the DEEP2 Galaxy Redshift Survey, we conclude that
among the environment estimators tested the projected $n^{\rm
th}$-nearest-neighbor distance measure provides the most accurate
estimate of local galaxy density over a continuous and broad range of
scales.
\end{abstract}

\keywords{methods:data analysis, methods:statistical,
galaxies:high-redshift, galaxies:statistics, surveys, large-scale
structure of universe}


\section{Introduction}

The observed properties of galaxies have long been known to depend
upon the environment in which they are located. For instance, red,
non-starforming galaxies (e.g. local ellipticals and lenticulars) are
systematically over-represented in highly over-dense environments such
as clusters \citep[e.g.][]{davis76,dressler80,postman84,balogh98}.
Recent studies have shown that the observed correlations between
galaxy properties and environment are not limited to the cores of rich
clusters, but extend to less dense domains including the outer regions
of clusters, galaxy groups, and the field
\citep[e.g.][]{balogh99, carlberg01, blanton03b, balogh04, croton05}. 

There are a variety of physical processes that can readily explain
these observational trends, including the action of dynamical
friction, tidal stripping, or gas pressure in dense
environments. These mechanisms, in combination with the hierarchical
model of galaxy formation \citep{kauffman93,somerville99,cole00}, in
which galaxies form in less dense environments and are then accreted
into larger groups and clusters, are generally consistent with the
current observations. From the empirical evidence, however, it remains
uncertain in what environment(s), by what mechanisms, and on what
time-scales galaxies evolve from a field-like population to a
cluster-like population. Still, the strong correlation of local
galaxy density with galaxy properties over a broad range of
environment does indicate that it plays an important role in galaxy
formation and evolution.

To study galaxy properties spanning a broad and continuous range of
local environment requires a thorough census of the 3-dimensional
galaxy distribution over a large volume. Such data sets are only
collected via large, systematic redshift surveys. For the nearby
galaxy population, wide-field spectroscopic and photometric surveys
(e.g.\ 2dFGRS, \citealt{colless01} and SDSS, \citealt{york00}) have
provided excellent data sets for studying galaxy environments ranging
from voids to rich clusters. Recent results from these large surveys
have found that galaxy environments correlate strongly with the
colors, luminosities, and morphologies of local galaxies
\citep[e.g.][]{balogh04,hogg03,blanton03b,hogg04}.

With the advent of new high-redshift surveys (e.g.\ the VLT-VIMOS Deep
Survey (VVDS), \citealt{lefevre03, lefevre04a} and the DEEP2 Galaxy
Redshift Survey (DEEP2), \citealt{davis03, faber05}), studies of
galaxy environment will be able to extend beyond the local
universe. Such deep, high-redshift surveys will provide a
representative snap-shot of the galaxy population and corresponding
local densities when the universe was half its present age, thereby
permitting an investigation into the influence of environment upon
galaxy evolution and formation. That is, extending environment studies
to higher redshift will enable a determination of whether the
correlations among galaxy properties observed in the local universe
are the result of physical processes acting over the entire lifetime
of the galaxy or whether the correlations were established during the
early formation of the galaxy.

While redshift surveys have grown in scale and studies of galaxy
environment have increased in prevalence, few published tests have
detailed the degree to which environment measures are affected by
survey limitations. For instance, the confined sky coverage of surveys
introduces geometric distortions (or edge effects) which bias local
density measures near boundaries or holes in the survey
field. Environment statistics can also be impacted by the redshift
precision and target selection requirements of a given survey. Mock
galaxy catalogs provide an excellent means for testing the biases
introduced to a given density measure by effects such as these.

In this paper, we test the applicability of several popular density
estimators within deep redshift surveys utilizing the mock galaxy
catalogs of \cite{yan04}. In particular, we investigate the effects of
redshift precision, survey field edges, redshift-space distortions,
and target selection. We also devote specific attention to the DEEP2
survey with the goal of identifying the optimal density measure for
use within the survey. Within this study, we do not consider global
measures of environment trends (such as correlation functions), but
instead focus on measures which can estimate environmental properties
of individual objects. The outline of the paper is as follows. In the
next section, we present the mock galaxy catalogs used to test
environment measures at high redshift. Subsequently (\S 3), we
describe the environment measures to be tested. In \S 4, we examine
the significance of redshift precision in determining local galaxy
densities. In \S 5, we conduct a detailed analysis of edge effects
with respect to each environment parameter. In \S 6 and \S 7, we then
address the influence of redshift-space distortions and target
selection on the various environment estimators. In \S 8, the roles of
completeness and the survey selection function are discussed. Finally
in \S9 and \S10, we conclude with a summary of the applicability of
each environment measure at high redshift and a discussion of the
suitability of current deep surveys to measuring local galaxy
densities.


\section{Simulating Deep Redshift Surveys}

Beginning with the Center for Astrophysics Redshift Survey
\citep{davis82}, large redshift surveys have played a major role in
studying galaxy properties, measuring cosmological parameters, and
studying the large-scale structure of the universe. With improvements
to astronomical instruments, local redshift surveys have ballooned in
size and surveys at high redshift $(z \sim 1)$ have become
possible. At present, high-$z$ surveys take two forms: (a) obtaining
precise $(\sigma_{z} \lesssim 0.001)$ redshifts using spectroscopic
observations of galaxies (e.g.\ DEEP2 and VVDS) and (b) using deep
photometry in many passbands to make less precise $(\sigma_z \gtrsim
0.05)$ photometric redshift estimates \citep[e.g.\
COMBO-17,][]{wolf03}. Each of these has its advantages and
disadvantages.

Even utilizing highly-multiplexed, multi-object spectrographs (e.g.\
DEIMOS, \citealt{faber03}) on large-aperture telescopes, a deep
spectroscopic redshift survey requires a vast amount of telescope time
and is invariably limited in the number of galaxies for which it can
measure redshifts. Slit or fiber collisions constrain the number of
objects able to be targeted during a given exposure while the forest
of OH sky lines in the optical and infrared plus instrument defects
and limited signal-to-noise cause redshifts to be missed for some
percentage of targeted objects. Spectroscopic surveys benefit from a
higher level of redshift precision which permits studies of kinematics
within galaxies and galaxy groups, while also measuring spectral
properties such as emission-line equivalent widths. On the other hand,
using an imager with a large field-of-view and observing in many
passbands, less precise photometric redshifts can be obtained
for nearly all galaxies above a given magnitude limit in the targeted
field. For this reason, photometric surveys are often able to build
larger samples and are ideal for measuring the galaxy luminosity
function or galaxy-galaxy lensing for which a high level of velocity
accuracy is not necessary.

In this paper, we employ the simulated galaxy catalogs of \cite{yan04}
to model both photometric and spectroscopic surveys at $z \sim 1$. The
simulations and all work in this paper employ a $\Lambda$CDM cosmology
with $\Omega_{M} = 0.3$, $\Omega_{\Lambda} = 0.7$, $h=1$, and
$\sigma_8 = 0.9$. The mock catalogs are derived from N-body
simulations by populating dark matter halos with galaxies according to
a halo occupation distribution (HOD) function
\citep{peacock00, seljak00} which describes the probability
distribution of the number of galaxies in a halo as a function of the
host halo mass. The luminosities of galaxies are then assigned
according to the conditional luminosity function (CLF) introduced by
\citet{yang03}, which describes the luminosity function in halos of
mass $M$. Models for the HOD and the CLF are constrained from the
2dFGRS luminosity function \citep{madgwick02} and two-point
correlation function \citep{madgwick03}. By assuming that the manner
in which dark matter halos are populated with galaxies does not evolve
from $z \sim 1$ to $z \sim 0$ \citep{yan03}, the mock catalogs are
built using simulation outputs at corresponding redshifts. The
simulated galaxy catalogs show excellent agreement with the lower-z
$(0.7 < z < 0.9)$ DEEP2 correlation function \citep{coil04a} and the
COMBO-17 luminosity function \citep{wolf03}; they therefore should
provide a realistic data set for studying measures of the environment
of galaxies at $z \sim 1$. For further details regarding the
construction of the mock catalogs, refer to \citet{yan03}.

From the volume-limited mock catalogs, we are able to mimic a typical
photometric redshift survey by selecting all galaxies above a given
magnitude limit and applying to each galaxy redshift a random offset
drawn from a Gaussian distribution with standard deviation
$\sigma_{z}$. We utilize the DEEP2 survey as a model high-redshift
spectroscopic survey. The volume-limited mock catalogs are selected
according to the DEEP2 magnitude limit of $R_{\rm AB} \le 24.1$ and
passed through the DEEP2 target-selection and slitmask-making code,
which is able to place approximately 60\% of available targets on
slitmasks for spectroscopy \citep{davis04}. Finally, 30\% of objects
are randomly rejected to reflect a conservative redshift success rate
of $\sim \! 70\%$. The 12 mock catalogs cover fields of $120' \times
30'$ in area with a total of $\sim \! 120$ DEEP2 slitmasks tiling the
1 square degree. To simulate larger survey fields, we tiled multiple
mock catalogs without overlap or discontinuity. Such large-field mocks
were essential for studying edge-effects (\S5) and for building large
sample sizes.

In each mock catalog, we have a complete tally of the total galaxy
distribution down to a luminosity of $0.1 {\rm L}_{*}$, along with
subsets of objects which pass the DEEP2 target-selection criteria,
were placed on a slitmask for observation, and yielded a successful
redshift. Such a census enables detailed study of the survey selection
function and the manner in which slitmask-making and target-selection
affect environment statistics. Throughout this paper, we utilize
several subsets drawn from the mock catalogs as described in Table
\ref{sample_tab}. Note that for each mock galaxy, the simulations
provide accurate positions in both real-space and redshift-space.

\begin{deluxetable}{l l}
\tablewidth{0pt}
\tablecolumns{2}
\tablecaption{\label{sample_tab} Subsamples selected from \\
the mock galaxy catalogs}
\tablehead{\colhead{Sample} & \colhead{Description}} 
\startdata
volume-limited & full mock galaxy catalog; $L > 0.1 L_{*}$ \\

magnitude-limited & $R_{\rm AB} \le 24.1$ \\

DEEP2-selected & $R_{\rm AB} \le 24.1$; \\
 & passed DEEP2 target-selection and \\
 & slitmask-making criteria; random $\sim \! 70\%$ \\
 & redshift success rate; high redshift \\
 & precision $(\sigma_z = 0.0001)$ \\

\enddata

\tablecomments{We present a list of commonly used subsamples \\
drawn from the mock galaxy catalogs of \citet{yan04}.}
\label{sample_tab}

\end{deluxetable}


\section{Environment Measures}

The environment of a galaxy is typically defined in terms of the
density of galaxies located in its immediate vicinity. However, a
variety of density measures are often employed in estimating
environment. For example, many previous analyses have focused on the
identification of predefined groups or clusters of galaxies, which can
be contrasted to those galaxies not inhabiting these over-dense
regions -- that is, the field population \citep[e.g.][]{kuntschner02,
vdb02, lewis02, christlein00}. Another approach is to instead derive a
continuous measure of the galaxy density distribution, such as by
measuring the distance to the $n^{\rm th}$-nearest neighbor
\citep[e.g.][]{gomez03, mateus04} or by directly smoothing the
observed galaxy distribution on a fixed scale \citep[e.g.][]{hogg03,
beuing02, kauffmann04}. The underlying theme in each of these methods
is that one requires a measure of the local number density of galaxies
at the position of each galaxy in the sample.

In our analysis, we focus on density estimators that do not rely on
identifying galaxy groups or clusters in any way. Lumping galaxies
into predefined classifications provides a poorly sampled range of
galaxy environments especially when compared to continuous measures of
environment. At high redshift, where dense regions are commonly
under-sampled and clusters and groups are less numerous, a more
continuous definition of environment is all the more desirable.
Still, identifying galaxies in groups at high $z$ is possible and has
been tested in a separate paper \citep{gerke05}. In this analysis, we
compare three popular density estimates: $n^{\rm th}$-nearest-neighbor
distance, counts in an aperture, and the Voronoi volume. This set of
measures is in no way presumed to be complete. Other promising methods
for measuring local galaxy density, including using a Gaussian kernel
to smooth the galaxy distribution over a given scale
\citep[e.g.][]{hogg03,balogh04}, are not discussed in this work.

\subsection{$n^{th}$-Nearest-Neighbor Distance, $D_{n}$
and $D_{p,n}$}

As first employed by \citet{dressler80}, the local galaxy density can
be estimated using the distance to the $n^{\rm th}$-nearest,
spectroscopically-observed neighbor of a given galaxy. Often, redshift
information is simply employed to exclude foreground and background
sources -- by restricting neighbors to a given velocity interval --
and the nearest-neighbor distance is measured in projection.
Commonly, the projected $n^{\rm th}$-nearest-neighbor distance,
$D_{p,n}$, is expressed as a surface density, $\Sigma_n = n / (\pi
D_{p,n}^2)$. Measuring nearest-neighbor distances in projection is
particularly useful when studying the density of galaxies in groups
and clusters \citep[e.g.][]{dressler80, lewis02}, where the
appropriate velocity interval by which to exclude background and
foreground galaxies may be selected according to the velocity
dispersion of the group or cluster. In this manner, one may
confidently exclude galaxies not associated with the group or cluster;
furthermore, as shown in \S 6, measuring in projection minimizes the
impact of redshift-space distortions.

For less dense environments or poorly sampled groups, there may be few
neighbors within the selected velocity interval, causing $D_{p,n}$ to
reflect the distance to other structures rather than the local
density. In environments where working in projection is problematic,
an alternative is to compute the $n^{\rm th}$-nearest-neighbor
distance in 3-dimensions by searching in spherical apertures for the
$n^{\rm th}$-nearest, spectroscopically-observed neighbor. Similar to
its projected counterpart, the 3-dimensional (3-d) $n^{\rm
th}$-nearest-neighbor distance, $D_{n}$, is often expressed as a
number density, $\rho_n = (3 n) / (4 \pi D_n^3)$.
Throughout this paper, all $n^{\rm th}$-nearest-neighbor distances are
quoted in units of comoving $h^{-1}$ Mpc and the symbols $D_{n}$ and
$D_{p,n}$ are employed to denote the 3-dimensional and projected
$n^{\rm th}$-nearest-neighbor distances, respectively.

\begin{figure}[h]
\plotone{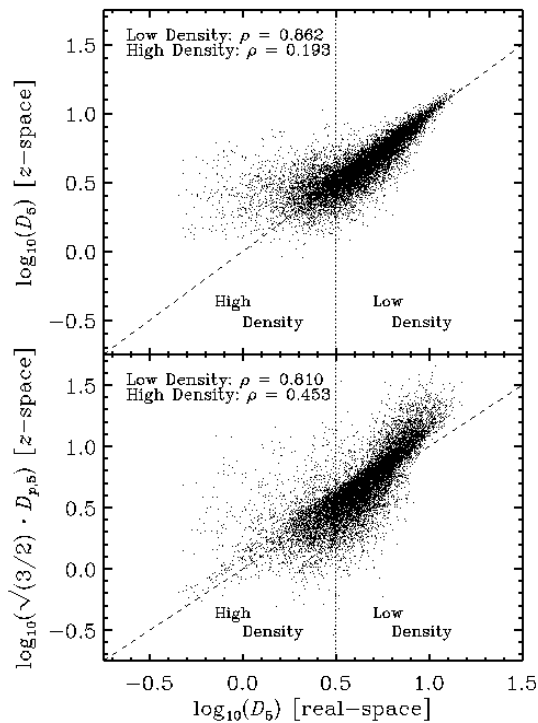}
\caption{(\emph{Top}) We plot $\log_{10}{(D_{5})}$ computed in 3-d
from the redshift-space distribution of galaxies in a $120' \times
60'$, DEEP2-selected mock catalog versus the 3-d $\log_{10}{(D_{5})}$
values measured using the real-space galaxy positions. (\emph{Bottom})
We plot projected $\log_{10}{(\sqrt{\frac{3}{2}} D_{p,5})}$ measured
from the redshift-space distribution of galaxies in the same
DEEP2-selected mock catalog versus the 3-d $\log_{10}{(D_{5})}$ values
measured from the real-space galaxy positions. For each galaxy, the
projected $n^{\rm th}$-nearest-neighbor distance was restricted to
neighbors within $\pm 1000\ {\rm km}/{\rm s}$ along the line-of-sight
and the distance has been scaled by a factor of $\sqrt{{3}/{2}}$ for
comparison to the 3-d measure of distance. The sample plotted consists
of 12,636 galaxies and is restricted to the redshift range $0.7 < z <
1.4$ and to galaxies more than $4\ h^{-1}$ comoving Mpc from the
nearest survey edge. For each sample, the Spearman ranked correlation
coefficient, $\rho$, is given for galaxies in the high density
$(\log_{10}{(D_{5})} \le 0.5)$ and the low density
$(\log_{10}{(D_{5})} > 0.5)$ regimes as defined by the real-space
$D_5$ values where the high density domain is selected to include the
most dense third of the sample. The dashed line in each plot follows a
correlation of $\rho = 1$.}
\label{nthnn}
\end{figure}

To study the effectiveness of the 3-d and projected $n^{\rm
th}$-nearest-neighbor distance measures at tracing the local density
of galaxies in different environments, we compute both $D_n$ and
$D_{p,n}$ for a DEEP2-selected sample consisting of 12,636 galaxies
as drawn from a $120' \times 60'$ mock catalog. In Figure
\ref{nthnn}, we compare the values of $D_n$ and $D_{p,n}$ for each
galaxy in the sample as measured using the redshift-space galaxy
positions to the measured value of $D_n$ as computed using the
real-space positions for each galaxy, which should reflect the true
local environment. We find that at high densities, where
redshift-space distortions are greater, the projected $n^{\rm
th}$-nearest-neighbor distance is superior at tracing the real-space
density of galaxies but still suffers greatly from peculiar
velocities. On scales corresponding to intermediate- and low-density
environments, the 3-d measure of $D_n$ is a slightly more accurate
estimate of the true galaxy distribution. For a DEEP2-selected mock
catalog, $\sim \! 15\%$ of the observed sample resides in the regime
$(\log_{10}(D_{5}) \le 0.5)$ where the 3-d $n^{\rm
th}$-nearest-neighbor distance saturates and loses sensitivity.

\begin{figure}[h]
\begin{center}
\plotone{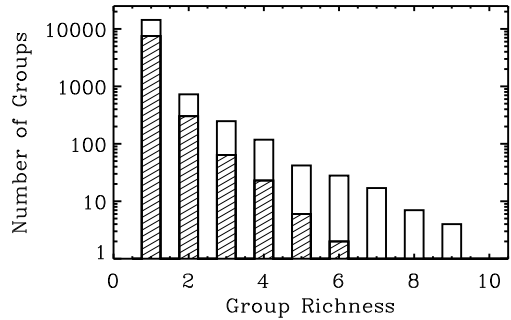}
\caption{We plot the distribution of group richness in a
magnitude-limited $(R_{\rm AB} \le 24.1)$, $120' \times 30'$ mock
catalog restricted to the redshift range $0.7 < z < 1.4$ -- \emph{open
boxes}. The \emph{crossed boxes} give the distribution in group
richness as observed in a DEEP2-selected sample. This observed group
catalog is derived from the magnitude-limited mock according to the
DEEP2 target-selection and slitmask-making processes described in
\S2. Each galaxy has been assigned to a group according to the
prescription of \citet{yan04}. Group richness is given by the number
of galaxies populating the group or virialized dark matter halo. Since
DEEP2 only samples the $z \sim 1$ galaxy population at a rate of $\sim
\! 45\%$, those groups with 7 or more members in the magnitude-limited
sample are typically observed to have a group richness of 3 or more in
the DEEP2 sample.}
\label{groups}
\end{center}
\end{figure}

In addition to a lack of sensitivity on given scales, the behavior of
the projected and the 3-d $n^{\rm th}$-nearest-neighbor distances
depends upon the choice of $n$. A measure of the $n^{\rm
th}$-nearest-neighbor distance effectively smoothes the galaxy
distribution in a non-linear fashion according to the adopted value of
$n$. If $n$ is chosen to be much larger than the richness of typical
groups in the sample, then the $n^{\rm th}$-nearest-neighbor distances
for galaxies in these groups will be pushed to erroneously high
values, as it will reflect the distance to the next-nearest
structure. In this work, we study both the projected and the 3-d
methods for computing the $n^{\rm th}$-nearest-neighbor distance
employing a variety of values for $n$. We limit most discussion to
values of $n = 2,3,5$ which correspond to the sizes of small groups
detected in the DEEP2 survey \citep{gerke05} and to the typical sizes
of groups in the mock catalogs (see Fig.\ \ref{groups}). For these
values of $n$, the sensitivity of the $n^{\rm th}$-nearest-neighbor
measure is rather independent of $n$. In under-dense environments, the
dependence on $n$ is very weak, while in dense environments the strong
clustering of groups \citep{padilla04, coil05} arranges to curb
$D_{n}$ for $n \gtrsim n_{\rm group}$. Also, in galaxy groups,
redshift-space distortions are a much greater source of error in $D_n$
than small variations in the choice of $n$.

For the projected $n^{\rm th}$-nearest-neighbor distance measure, we
test the sensitivity of $D_{p,n}$ using line-of-sight velocity
intervals ranging from $\pm 750\ {\rm km}/{\rm s}$ to $\pm 2000\ {\rm
km}/{\rm s}$. As shown in Table \ref{vint_tab}, using a larger
velocity interval by which to exclude foreground and background
sources increases the accuracy of $D_{p,n}$ in dense environments but
also sacrifices sensitivity at low densities. We find that for a
DEEP2-selected sample, a velocity interval of $\pm1000\ {\rm km}/{\rm
s} - \pm1500\ {\rm km}/{\rm s}$ is best suited for a broad range of
environments. Compared to photometric redshift errors (in the best
datasets, $\sigma_z \sim 6000\ {\rm km}/{\rm s}$), the sizes of the
tested line-of-sight velocity windows are small. However, larger
velocity intervals sacrifice sensitivity on small scales and provide
poorer measures of the local density about each galaxy; a window large
enough not to be dominated by photometric redshift errors is also
large compared to the typical length-scales of large-scale structure
(e.g. the correlation length and typical void sizes), and thus
provides a poor measure of environment.

\begin{deluxetable}{c c c c c c}
\tablewidth{0pt}
\tablecolumns{6}
\tablecaption{\label{vint_tab} Dependence of $D_{p,n}$ on velocity window}
\tablehead{ & $n$ & $\Delta v\ ({\rm km}/{\rm s})$ & $\rho_{\rm high}$ &
      $\rho_{\rm low}$ & }
\startdata
&  3   &       750            &     0.454     &   0.786 &   \\
&  3   &      1000            &     0.540     &   0.744 &   \\
&  3   &      1500            &     0.598     &   0.659 &   \\
&  3   &      2000            &     0.600     &   0.602 &   \\
\hline
&  5   &       750            &     0.384     &   0.831 &   \\
&  5   &      1000            &     0.451     &   0.808 &   \\
&  5   &      1500            &     0.496     &   0.736 &   \\
&  5   &      2000            &     0.505     &   0.673 &   \\
\enddata
\tablecomments{For a range of velocity intervals $(\Delta v)$ from $\pm
750\ {\rm km}/{\rm s}$ to $\pm 2000\ {\rm km}/{\rm s}$, we compute the
projected $n^{\rm th}$-nearest-neighbor distance for the galaxies in a
$120' \times 60'$ DEEP2-selected mock catalog where $\Delta v$ is the
velocity window over which to exclude foreground and background
interlopers. Each set of $D_{p,n}$ values are correlated against the
real-space $D_{n}$ distances computed for the same mock catalog, as
illustrated in Figure \ref{nthnn}. We present the Spearman ranked
correlation coefficients, $\rho$, in the low density $(\rho_{\rm
low})$ and high density $(\rho_{\rm high})$ regimes. For $n=5$, the
division between the low and high density regimes is made at
$\log_{10}{(D_5)} = 0.5$, while for $n=3$ the distinction is drawn at
$\log_{10}{(D_3)} = 0.4$. }
\label{vint_tab}
\end{deluxetable}

\subsection{Counts in an Aperture, $C$}

Another method for estimating the local galaxy density is to count
galaxies within a fixed metric aperture. For example, \cite{hogg03}
count spectroscopically-observed galaxies in the SDSS within spheres
of radii $8\ h^{-1}$ comoving Mpc centered on each
spectroscopically-observed galaxy. In high-redshift surveys where the
survey field may cover $\sim \! 1$ square degree or less, such a large
spherical aperture will be dramatically affected by the survey edges
(see \S5). For instance, within a $30'' \times 30''$ field (i.e. $20\
h^{-1}$ comoving Mpc on a side at $z \sim 1$), 81\% of spherical
apertures with a radius of $1\ h^{-1}$ comoving Mpc will fit within
the surveyed field, while for apertures of radius $3\ h^{-1}$ and $5\
h^{-1}$ comoving Mpc only 49\% and 25\% of the field, respectively,
will be unaffected by edges. Furthermore, local studies indicate that
larger apertures do not provide any advantage or additional
information worth this high price. Both observational and theoretical
studies, suggest that galaxy properties are more closely related to
dark matter halo mass and small-scale environment than the large-scale
environment of the galaxy \citep[e.g.][]{lemson99,blanton04}.

While choosing smaller spherical apertures would reduce the amount of
survey volume affected by edges, apertures smaller than approximately
$\pm 1000\ {\rm km}/{\rm s}$ along the line-of-sight will not be
sensitive to galaxy groups or clusters. The counts in an aperture
measure effectively smooths the data on some adopted scale, thereby
losing sensitivity on smaller and larger scales. In our analysis, we
employ a series of cylindrical apertures measuring 1--2 $h^{-1}$
comoving Mpc transverse (radius) and $\pm 500\ {\rm km}/{\rm s} - \pm
2000\ {\rm km}/{\rm s}$ along the line-of-sight. The dimensions of our
cylindrical apertures are chosen to match the typical sizes of halos
in the simulations \citep{yan04}.

\subsection{The Voronoi Volume, $V$}
The Voronoi volume is a geometric measure that has seen use from
engineering and biology to astronomy \citep{ramella01,
marinoni02}. Unlike counts in an aperture, the Voronoi volume does not
smooth the galaxy distribution in any way. It provides a continuous,
adaptive measure of galaxy density on all scales by measuring a unique
volume about each spectroscopically-observed galaxy.

\begin{figure}[h]
\begin{center}
\plotone{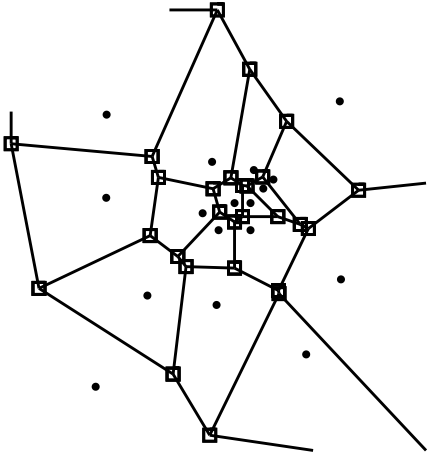}
\caption{The two-dimensional Voronoi diagram (Dirichlet tessellation)
for an array of points (that is, a distribution of galaxies) is
illustrated. The points represent the galaxies and the open squares
identify the Voronoi vertices. Note that the areas of the polygons
depend strongly on the local density of points. Similarly,
three-dimensional Voronoi cell volumes vary inversely with local
density.}
\label{voronoi}
\end{center}
\end{figure}

As illustrated in Figure \ref{voronoi}, the Voronoi partition of space
is the three-dimensional analogue of the two-dimensional Dirichlet
tessellation, in which a plane containing a set of data points is
divided into a set of polygons, each containing one of the points. A
Voronoi polyhedron is the unique three-dimensional convex region of
space surrounding a data point (the seed), such that within the
polyhedron every point is closer to the seed than to any other data
point. The faces of the Voronoi polyhedron are defined by the
perpendicular bisecting planes of the vectors connecting the seed to
its neighbors, where a seed's neighbors are those points connected to
it by the Delaunay complex -- the set of tetrahedra whose vertices are
at the data points and whose unique, circumscribing spheres contain no
other data points. The Voronoi partition and Delaunay complex are thus
geometrical duals of one another.

Computing the Voronoi partition for a galaxy redshift survey provides
a natural way to measure the local density of galaxies, since the
volume of a galaxy's Voronoi polyhedron will vary inversely with the
distance to its closest neighbors. For this reason, the Voronoi volume
associated with each galaxy serves as a natural parameterization of
that galaxy's environment. Galaxies in dense regions will have small
Voronoi volumes, while isolated galaxies will have larger volumes. Our
methods for computing the Voronoi partition are identical to those of
\cite{marinoni02}, and we refer the reader to that work for
computational details and for further discussion of the usefulness and
historical context of the Voronoi partition and Delaunay complex. In
this paper, we will employ the symbol $V$ to denote the Voronoi volume
of a given galaxy and all Voronoi volumes are measured in units of
comoving $(h^{-1}\ {\rm Mpc})^{3}$.

\section{Redshift Precision and Target Selection Rate: 
Photometric versus Spectroscopic Surveys}

As discussed in \S 2, photometric and spectroscopic redshift surveys
differ in the precision with which they are able to measure galaxy
redshifts. To test the significance of redshift precision in measuring
local galaxy environment, we have produced a variety of mock surveys
with differing characteristics. First, we simulate two photometric
redshift surveys which mimic the varying precisions of the COMBO-17
photometric redshift survey quoted in the literature. Our first
simulated photometric redshift survey adopts a magnitude limit of
$R_{\rm AB} \le 24.1$ and redshift uncertainty of $\sigma_z \approx
0.02$ which reaches equally deep and is more precise than the COMBO-17
specifications of $R_{vega} \lesssim 24$, $\sigma_z \approx 0.03$ as
given by \citet{wolf03}. In addition, we simulate a photometric
redshift survey with the same magnitude limit of $R_{\rm AB} \le 24.1$
and a redshift precision of $\sigma_z \approx 0.05$ as specified for
COMBO-17 by \citet{taylor04}. Both magnitude-limited samples are drawn
from the same volume-limited mock catalog and include 22,961 galaxies
covering a $120' \times 30'$ field. Note that our assumed redshift
uncertainties are lower limits to the redshift precision for the
COMBO-17 survey. As discussed by \citet{bell04}, the photometric
redshift precision for COMBO-17 depends strongly on the galaxy type and
redshift; galaxies such as starbursts which lack a strong
4000\AA-break yield redshifts with much greater uncertainties
$(\sigma_{z} \sim 0.1)$, while at higher redshifts K-correction
uncertainties introduce systematic redshift errors.

Running the same volume-limited galaxy catalog through the DEEP2
target-selection and slitmask-making software and assuming a
conservative redshift success rate (see \S 2), we also produce a mock
spectroscopic sample with redshift precision of $\sigma_{z} \sim
0.0001$, mimicing the DEEP2 redshift survey. This DEEP2-selected
spectroscopic sample includes 9,302 galaxies covering the same 1
square degree field (sampling $\sim \! 50\%$ of galaxies to the
magnitude limit). To simulate the VVDS ``deep survey'' in the CDF-S
\citep{lefevre04a}, we randomly select 25\% of objects to the same
$R_{\rm AB} \le 24.1$ magnitude limit. This ``VVDS-selected'' sample
is an optimistic simulation of VVDS, assuming a 100\% redshift success
rate \citep{vanzella04} and ignoring differences in the bandpass
used. Lastly, as a comparison sample, we select the full
magnitude-limited mock (22,961 galaxies at $R_{\rm AB} \le 24.1$)
assigning redshifts according to the real-space positions of the
galaxies as defined in the mock simulations. Each environment
estimator ($n^{\rm th}$-nearest-neighbor distance, Voronoi volume, and
counts in an aperture) is then computed on the photometric,
spectroscopic, and real-space galaxy samples.

For this comparison, we restrict our analysis to the redshift range
$0.7 < z < 1.4$ and to only those galaxies at transverse distances of
greater than $4\ h^{-1}$ comoving Mpc from the nearest edge in the
survey volume. These restrictions make edge effects in both the
redshift and transverse directions negligible, but do not introduce
any selection biases (see \S5). Note that the mock catalogs are not
subject to interior edges; that is, the simulations cover a contiguous
1 square degree of sky with no holes.

\begin{figure*}[t!]
\begin{center}
\plotone{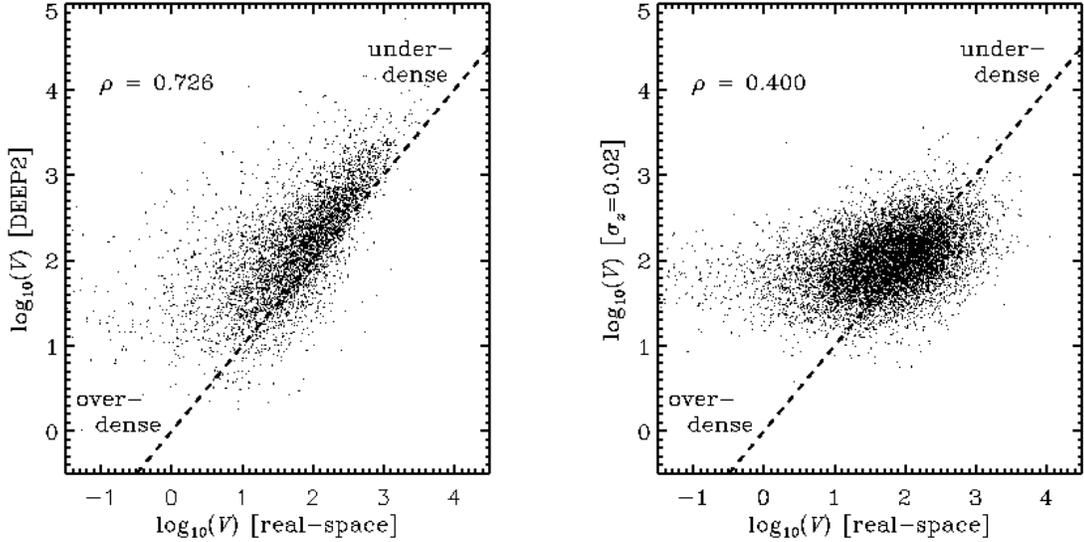}
\caption{(\emph{Left}) We plot a comparison of Voronoi volumes
measured for the 4,539 galaxies in a $120' \times 30'$, simulated
DEEP2 spectroscopic redshift survey sample $(\sigma_z \sim 0.0001)$
which are more than $4\ h^{-1}$ comoving Mpc away from the nearest
survey edge and within the redshift range $0.7 < z < 1.4$. Plotted is
the log of the Voronoi volumes as measured in the simulated
DEEP2-selected sample versus the log of the Voronoi volumes as
measured on the full magnitude-limited, real-space
catalog. (\emph{Right}) We plot a comparison of Voronoi volumes
measured for the 9,310 galaxies in a high-precision $(\sigma_z \sim
0.02)$ simulated photometric redshift survey sample which were more
than $4\ h^{-1}$ comoving Mpc away from the nearest survey edge and
within the redshift range $0.7 < z < 1.4$. Plotted is the log of the
Voronoi volumes as measured in the simulated photometric survey sample
versus the log of the Voronoi volumes as measured on the full
magnitude-limited, real-space catalog. In each plot, the Spearman
ranked correlation coefficient, $\rho$, is given in the upper-left
corner. The dashed lines follow a correlation of $\rho = 1$. While
both samples are influenced by redshift-space distortions (see \S 6)
and the spectroscopic sample (\emph{left}) suffers from poorer
sampling, the inferior redshift precision of the photometric survey
causes the line-of-sight galaxy distribution to smeared out
irretrievably on all but the largest scales.} 
\label{zaccuracy}
\end{center}
\end{figure*}

We find that the precision of even the best photometric redshifts is
not sufficient to measure local galaxy environments. Figure
\ref{zaccuracy} shows the comparison between Voronoi volumes, $V$, as
measured using the real-space galaxy positions compared to those
calculated using the observed redshifts for two representative
surveys. Even assuming redshift errors as small as $\sigma_z \approx
0.02$, the environment measured in a photometric redshift survey is
insensitive for all but the very lowest density environments; the
Spearman ranked correlation coefficient between the real-space and
photometric measures of Voronoi volumes is $\rho = 0.4$. For the
spectroscopic survey, redshift-space distortions introduce some
scatter at high densities, but the overall distribution of
environments is well measured. In all, the Voronoi volumes measured
from the observed spectroscopic redshift distribution trace the
real-space Voronoi volumes with much greater precision, yielding a
correlation coefficient of $\rho = 0.73$. Very similar results are
observed for the $D_n$, $D_{p,n}$, and $C$ environment estimators.

In Table \ref{zacc_tab}, we expand our analysis to a better sampled
range of redshift uncertainties. Even if the precision of photometric
redshifts is greatly improved -- by a factor of 2 or 4 -- we find that
low-resolution spectroscopic surveys with galaxy sampling comparable
to DEEP2 provide a significantly better trace of the 3-dimensional
galaxy environment. It is only at very high redshift precisions
$(\sigma_z \lesssim 0.005)$ and when measuring densities in projection
that photometric redshift surveys are able to rival their
spectroscopic counterparts as probes of galaxy environment.

Among the spectroscopic redshift surveys simulated, the greater
sampling and improved redshift precision of the DEEP2 survey prove
significantly superior to the VVDS in tracing the real-space density
of galaxies. On the other hand, at precisions better than the $30\
{\rm km}/{\rm s}$ uncertainty in DEEP2 redshifts, redshift-space
distortions dominate the ability to measure local densities and
thereby limit any advantage of improved redshift measurements (see
Table \ref{zacc_tab}). Note that the galaxy samples in Table
\ref{zacc_tab} transition from a magnitude-limited $(R_{\rm AB} \le
24.1)$ sample at low redshift precision to mimic photometric or grism
spectroscopic redshift surveys to a sample selected using the DEEP2
target-selection and slitmask-making procedures or a VVDS-like
selection to simulate higher-resolution spectroscopic surveys, which
have superior redshift precision but lower sampling rates.

\begin{deluxetable}{c c c c c}
\tablewidth{0pt}
\tablecolumns{5}
\tablecaption{\label{zacc_tab} Environment measurements in
photometric and spectroscopic surveys}
\tablehead{Sample & $\sigma_z$ & $\rho_V$ & $\rho_{D_5}$ &
$\rho_{D_{p,5}}$ }
\startdata
$R_{\rm AB} \le 24.1$ & 0.05   & 0.307 & 0.310 & 0.389 \\
\boldmath$R_{\rm AB} \le 24.1$ & {\bf 0.02}   & {\bf 0.400} & {\bf
0.396} & {\bf 0.473} \\
$R_{\rm AB} \le 24.1$ & 0.01   & 0.494 & 0.478 & 0.575 \\
$R_{\rm AB} \le 24.1$ & 0.005  & 0.596 & 0.579 & 0.688 \\
$R_{\rm AB} \le 24.1$ & 0.0025 & 0.675 & 0.677 & 0.803 \\
{\bf VVDS-selected} & {\bf 0.001} & {\bf 0.640} & {\bf 0.655} & {\bf 0.689} \\
DEEP2-selected        & 0.0025 & 0.611 & 0.625 & 0.704 \\
DEEP2-selected        & 0.001  & 0.691 & 0.716 & 0.785 \\
{\bf DEEP2-selected}  & {\bf 0.0001} & {\bf 0.726} & {\bf 0.749} & {\bf 0.802} \\
DEEP2-selected        & 0      & 0.726 & 0.751 & 0.801 \\
\enddata
\tablecomments{For a range of redshift precisions $(\sigma_z)$ 
we compute the $5^{\rm th}$-nearest-neighbor distance $(D_5)$, Voronoi
volume $(V)$, and the projected $5^{\rm th}$-nearest-neighbor distance
$(D_{p,5})$, for the galaxies in a $120' \times 30'$ mock catalog. As
described in the main text, three samples are selected from the same
mock catalog according to ({\it a}) the DEEP2 target-selection and
slitmask-making procedure [DEEP2-selected], ({\it b}) a magnitude
limit of $R_{\rm AB} \le 24.1$, and ({\it c}) a 25\% random sampling
to the same magnitude limit of $R_{\rm AB} \le 24.1$
[VVDS-selected]. Restricting to the redshift range $0.7 < z < 1.4$ and
more than $4\ h^{-1}$ comoving Mpc removed from a field edge, the
magnitude-limited sample contains 9,310 galaxies while the
DEEP2-selected and VVDS-selected samples include 4,538 and 2,375 of
those galaxies, respectively. For each sample and redshift precision,
we (rank) correlate estimates of $V$, $D_{5}$, and $D_{p,5}$ with
similar measures computed on a magnitude-limited sample using the
real-space galaxy positions. Note that $\sigma_z = 0.0001$ corresponds
to the redshift precision of the DEEP2 survey and that surveys
reflecting the COMBO-17, VVDS, and DEEP2 attributes have been
highlighted in bold font in the table.}
\label{zacc_tab}
\end{deluxetable}


\section{Edge Effects} 

When measuring galaxy densities within any survey, one must always be
careful of edge effects introduced by the limited area of sky covered
in the survey. Even using the largest optical telescopes and an
instrument with a generous field-of-view, a deep redshift survey is
limited in its ability to cover large regions. Furthermore, to
minimize the effects of cosmic variance on the data set, a
high-redshift survey is likely to spread the sky coverage over several
fields. This limits the amount of contiguous sky coverage and
increases the proportion of the survey area that is near an edge. In
addition to the edges created by the chosen geometry of the survey
field(s), edges and holes can be created in the data set by effects
such as bright stars in a field or problematic regions in photometric
detection which prohibit any galaxies from being targeted there. To
start, we will restrict our discussion to survey edges in the plane of
the sky, but later discussion will address edges in the line-of-sight
direction.

The general effect of edges on each density estimator is to push the
measurement towards lower density. To quantify the degree to which
each environment measure is affected by edges we compute each measure
on a large DEEP2-selected spectroscopic mock galaxy catalog covering a
wide field, $120' \times 90'$. From the center of this larger
simulation we extract a smaller rectangular survey field, covering
$40' \times 30'$, and rerun each environment measure on this data
set. For every galaxy in the smaller survey field, we then have
measurements of environment unaffected by edges (when measured on the
larger sample) and measurements in which survey edges play a greater
role (when measured on the smaller field). Trimming to the redshift
range $0.7 < z < 1.4$, the smaller sample consists of 2,803
galaxies. In the following subsections, we discuss how each
environment measure is affected by the edges of the survey region on
the plane of the sky. Note that we also ran tests incorporating holes
and irregular survey edges with very little change in the relative
results for the tested environment measures.

For each galaxy in the studied sample, we express the difference in a
given environment measure due to survey edges as a fraction of the
width of the distribution for that measure. Specifically, we define
the percent change in environment measure $X$ for galaxy $q$ by
\begin{equation}
  \Delta_{e}(X) = \frac{ \log_{10}{\left(X_{2,q}\right)} -
	\log_{10}{\left(X_{1,q}\right)} }{ \sigma_{1} } \cdot 100\%
\label{eq_edge}
\end{equation}
where $X_{2,q}$ is the measure of $X$ for galaxy $q$ computed on the
smaller mock, $X_{1,q}$ is similarly computed on the wide-field mock,
and $\sigma_{1}$ is a measure of the Gaussian width of the logarithmic
distribution of environment measure $X$ as calculated in the larger
simulation. Quantifying the change in each environment measure in this
fashion enables the role of edge effects to be compared between
different environment estimators in a uniform manner.

\subsection{Survey Edges and $n^{th}$-Nearest-Neighbor
Distance} 

The $n^{\rm th}$-nearest-neighbor distance environment measure -- in
both projection and 3-dimensions -- is affected by edges in a
predictable manner. Any galaxy with an edge located closer than the
measured $n^{\rm th}$-nearest-neighbor distance must be affected by
the survey edges. However, to remove all such galaxies based on this
criterion ($D_{n} > D_{\rm edge}$ or $D_{p,n} > D_{\rm edge}$) biases
the sample towards over-dense environments by excluding less-dense
regions over a greater volume than more-dense regions. To avoid such
biasing of the sample, a simple cut in edge distance can be made --
excluding all galaxies within some distance of the nearest edge. This
cut introduces no environment-dependent bias, but does allow some
contamination to the sample at under-dense environments depending on
the severity of the cut. In our simulations, we find that removing all
galaxies within $2\ h^{-1}$ comoving Mpc of a survey edge creates a
catalog with minimal contamination (roughly 5\% of the sample has
$\Delta_{e}(D_3) > 10\%$) and still retains 65\% of the data
set. Relaxing the constraint to $D_{\rm edge} > 1\ h^{-1}$ comoving
Mpc, the level of contamination in the sample doubles to $\sim \!
10\%$ with $\Delta_{e}(D_3) > 10\%$ while the percentage of the sample
retained increases to $\sim \!  85\%$.

\begin{figure}[h]
\begin{center}
\plotone{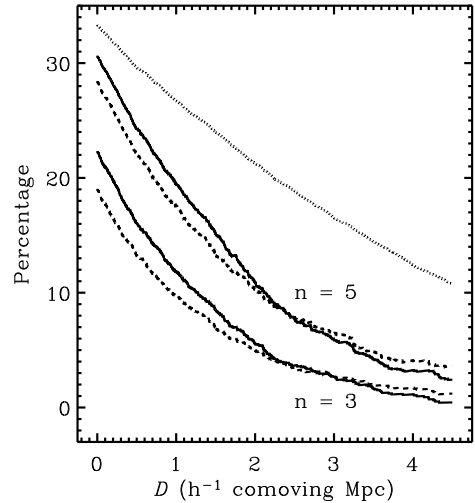}
\caption{For the smaller sample of 2,803 galaxies, we plot the
percentage of the simulated sample with $D_{\rm edge} > D$
(\emph{dotted line}) where $D_{\rm edge}$ is the projected distant
from a given galaxy to the nearest survey edge. Note that the dotted
line has been scaled down by a factor of 3 in order to fit within the
plotting range. Of those galaxies meeting the edge-distance condition
$(D_{\rm edge} > D)$, we also plot the percentage of galaxies
corrupted by a survey edge on the plane of the sky according to the
condition, $\Delta_{e}(X) > 10\%$. For $n=3$ and $n=5$, the
\emph{dashed lines} and \emph{solid lines} illustrate the
contamination of $D_{p,n}$ and $D_n$, respectively, as a function of
the edge restriction $D_{\rm edge} > D$. In measuring $D_{p,n}$, we
employ a velocity interval of $\pm 1000\ {\rm km}/{\rm s}$.}
\label{pcent_plot}
\end{center}
\end{figure}

As illustrated in Figure \ref{pcent_plot}, we find that edge-effects
show a clear dependence on $n$; the level of contamination due to
survey edges in the plane of the sky increases by roughly a factor of
two for $D_{5}$ relative to $D_{3}$. Also, for a fixed value of $n$,
the projected $n^{\rm th}$-nearest-neighbor distance is slightly more
robust to edge-effects in the regime where sample size is maximized
($D < 2\ h^{-1}\ {\rm Mpc}$ in Figure \ref{pcent_plot}).

\subsection{Survey Edges and Counts in an Aperture}
For an aperture of fixed comoving size, the edge effects upon the
counts in an aperture density measurement are easily understood and
cleaned from the sample. Only galaxies located within $r_{\rm t}$ of
an edge are affected, where $2 r_{\rm t}$ is the transverse diameter of
the chosen aperture. Thus, by removing any galaxies within $r_{\rm t}$
of an edge, the sample is entirely devoid of edge-affected
galaxies. Such a trimming of the data set does not introduce a
selection effect, that is, there is no bias towards environments of a
given sort. 

In our simulated spectroscopic data set of 2,803 galaxies, $\sim \!
15\%$ of the sample are positioned within $r_{\rm t}$ of a survey edge
using a cyclindrical aperture of $r_{\rm t} = 1\ h^{-1}$ comoving
Mpc. However, only for a scant $\sim \! 3\%$ of the sample did we find
$\Delta C \neq 0$ when comparing measurements of $C$ made on the
smaller simulation to those made on the wide-field sample. One
possible means for salvaging some edge-affected galaxies would be to
scale the measured counts in each aperture by the amount of the
aperture contained within the survey area. Due to the low rate at
which $C$ is perturbed by survey edges, however, such a correction
actually causes an overestimate of $C$ for the majority of galaxies
near the edge $(D_{\rm edge} < r_{\rm t})$ of the survey field.

\subsection{Survey Edges and Voronoi Volumes}
Due to the geometrical complexity of the Voronoi tesselation,
understanding the effects of edges on the calculated Voronoi volumes
is less straightforward than for the previously discussed density
measures. For galaxies very close to exterior edges in the survey
field, Voronoi volumes can be unbounded and such galaxies should be
consequently discarded from the sample. On a more subtle level, edge
effects -- including interior edges -- will also cause volumes to be
increased in size while the volumes still remain bounded. Some of
these edge-affected volumes can be detected as having Voronoi vertices
outside of the survey field. However, many other edge-affected Voronoi
volumes are not detectable in such a manner. 

\begin{figure}[h]
\begin{center}
\plotone{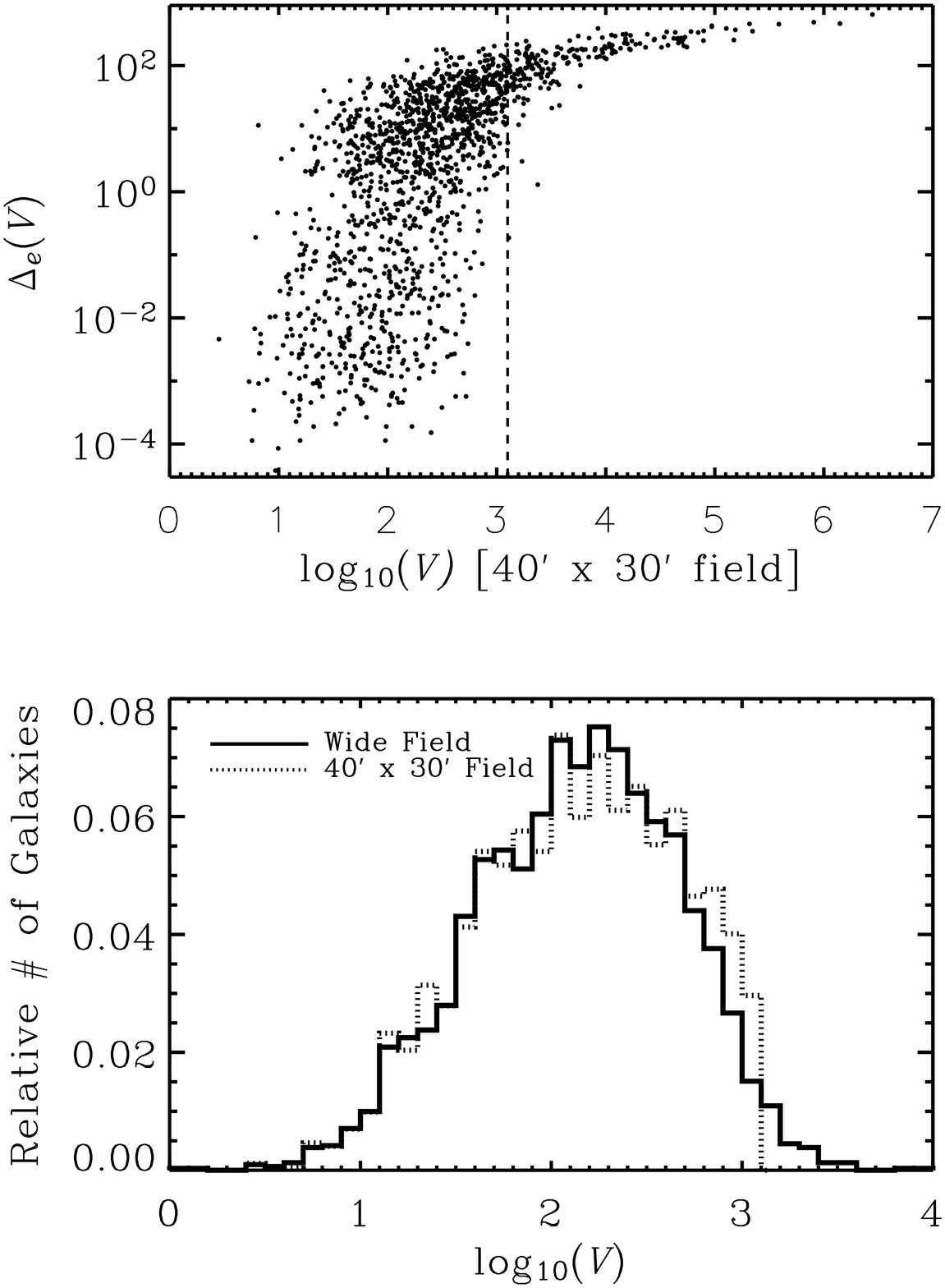}
\caption{(\emph{Top}) For the 1,790 galaxies located more than $2\ h^{-1}$
comoving Mpc from an edge, we plot $\Delta_{e}(V)$ versus the log of
the Voronoi volume, $V$, as measured in the smaller field. Note that
$\Delta_{e}(V)$ is given by equation \ref{eq_edge}. The clear trend is
for galaxies in large volumes to be more severely affected by survey
edges. The dashed line shows a proposed cut in Voronoi volume
$(\log_{10}{(V)} < 3.1)$ designed to remove highly edge-affected
galaxies from the sample. (\emph{Bottom}) We plot the distribution of
Voronoi volumes for the full sample of 2,803 within the redshift
window $0.7 < z < 1.4$ as measured on the wide-field mock (solid) and
for the sample of 1,720 galaxies greater than $2\ h^{-1}$ comoving Mpc
of an edge and with $\log_{10}{(V)} < 3.1$ as measured on the smaller
simulation (dotted). Note each histogram has been normalized to have
an integral of unity. The overall distributions are in good agreement
up to $\log_{10}{(V)} = 2.7$. The excess of galaxies with volumes $2.7
< \log_{10}{(V)} < 3.1$ as measured in the smaller simulation results
from edge-affected galaxies contaminating the final sample.}
\label{edge1}
\end{center}
\end{figure}

As illustrated in Figure \ref{edge1}, even excluding galaxies located
near a survey edge (e.g.\ within $2\ h^{-1}$ comoving Mpc), the
distribution of Voronoi volumes is still greatly affected by edges
with a bias towards large volumes being pushed to even larger values
(see Fig.\ \ref{edge1}). It is possible to minimize this effect by
retaining only galaxies with $V$ below some limit. In our simulations,
by truncating at $\log_{10}(V) = 3.1$, the amount of contamination due
to edge effects can be reduced to approximately 20\% of the sample
with $\Delta_{e}(V) > 10\%$. While making such cuts according to
distance to the nearest edge $(D_{\rm edge} > 2\ h^{-1}\ {\rm
comoving}\ {\rm Mpc})$ and Voronoi volume $(\log_{10}(V) < 3.1)$
effectively reduces the number of edge-affected galaxies in the
sample, it also restricts the dynamic range of the Voronoi measure and
considerably reduces the size of the sample. For our simulated
spectroscopic sample of 2,803 galaxies, the Voronoi volume density
measure was the most dramatically affected by edges with $\gtrsim \!
45\%$ of the sample being corrupted, that is, having $\Delta_{e}(V) >
10\%$.

\subsection{Effects of Finite Redshift Range}

As a secondary effect, the finite redshift range probed by any survey
imposes edges in the line-of-sight direction. The role of these edges
is more easily handled by restraining all scientific analysis to a
limited, well-sampled range of redshifts. In the DEEP2 survey, the
ability to measure redshifts at $z > 1.4$ or $z < 0.7$ decreases
significantly as the [OII] emission-line doublet leaves the observed
optical window. In a DEEP2-selected mock catalog, we find that by
restricting our sample to those galaxies at $0.7 < z < 1.4$, less than
1\% of the sample has a $5^{\rm th}$-neareast-neighbor distance,
$D_{5}$, greater than the distance to the $z=0.7$ or $z=1.4$
edge. Similar contamination rates are found for the other environment
estimators.

A second concern for spectroscopic redshift surveys is the possibility
of missing redshifts over specific wavelength intervals, especially in
the far-optical and near-infrared where OH sky lines can dominate the
spectrum. At $z \sim 1$ where optical surveys often rely upon a
singular spectral feature (e.g.\ the [OII] doublet at $\lambda_{\rm
rest} \sim 3727{\rm \AA}$) for redshift measurements, a hole in
wavelength coverage translates directly into a hole in the survey's
redshift sampling. For the DEEP2 survey, the high-resolution $(R \sim
5000)$ of the DEIMOS data minimizes this effect, as the sky lines are
then narrower than the components and spacing of the [OII] doublet. In
truth, the DEEP2 redshift distribution exhibits no significant
cross-correlation with a sky spectrum mapped to redshift according to
either the central wavelength of the [OII] doublet or the wavelengths
of either subcomponent \citep{newman05b}. However, for lower
resolution surveys such as the VVDS, windows of redshift insensitivity
may be a concern that must be addressed in measuring galaxy densities.


\section{Redshift-Space Distortions} 

While spectroscopic redshift surveys are able to measure galaxy
redshifts with great precision, redshift measurements by nature are
measurements of velocity and not distance. Accordingly, converting
differences in redshift to relative line-of-sight distances is subject
to the peculiar velocities of the galaxies. Such peculiar motions are
greatest in dense regions such as groups or clusters where the
velocity dispersion of the group causes the inter-member spacing to be
larger in redshift space than in real space. Due to this environmental
dependency of redshift-space distortions, it is essential to
understand the manner in which they affect a given galaxy density
measure.

Within a mock DEEP2-selected spectroscopic galaxy catalog covering
$120' \times 90'$, we compute each environment measure using the both
real-space positions of the galaxies and the observed redshift-derived
positions. Restraining our analysis to galaxies at edge distances
greater than $4\ h^{-1}$ comoving Mpc and within the redshift range
$0.7 < z < 1.4$, we quantify the effect of redshift-space distortions
on each environment estimator by calculating the change in each
measure as computed on the real-space mock relative to the
corresponding measure derived from the observed spectroscopic mock. As
in \S5, we express the difference in a given environment measure as a
fraction of the width of the real-space distribution for that
environment measure. For example, the percent change in environment
measure $X$ for galaxy $q$ is given by
\begin{equation}
\Delta_{z}(X) = \frac{
	\log_{10}{\left(X_{z,q}\right)} -
	\log_{10}{\left(X_{R,q}\right)} }{ \sigma_{R} } \cdot 100\%
\label{zseqn}
\end{equation}
where $X_{z,q}$ is the measure of $X$ for galaxy $q$ computed from the
redshift-derived position and $X_{R,q}$ is similarly computed from the
real-space position. The width, $\sigma_{R}$, is determined via a
Gaussian fit to the logarithmic distribution of environment measure
$X$ for all galaxies in the real-space simulation. Here, $\sigma_{R}$
can be measured on the real-space distribution of $\log_{10}{(X)}$ or
the redshift-space distribution with negligible difference for the
DEEP2-selected sample.

\begin{figure}[h]
\begin{center}
\plotone{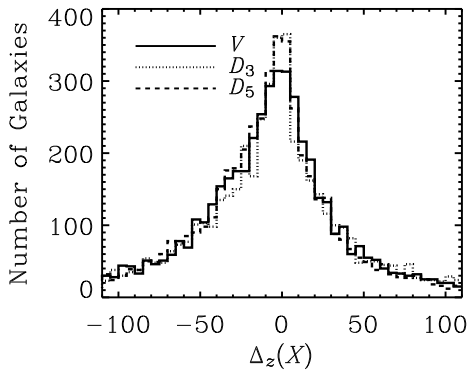}
\caption{The effect of redshift-space distortions on the $V$ and $D_n$
environment measures is illustrated. We plot the distribution of
$\Delta_{z}(X)$ (see eq.\ \ref{zseqn}) for the 5,031 galaxies in
the DEEP2-selected simulation within $0.7 < z < 1.4$ and more than $4\
h^{-1}$ comoving Mpc from a survey edge. The Voronoi volume and $D_n$
statistic are similarly affected by redshift-space
distortions. Working in projection, for example using $D_{p,n}$ or
$C$, provides a much more robust estimate of galaxy density in
over-dense environments.}
\label{zspace1}
\end{center}
\end{figure}

\begin{figure}[h]
\begin{center}
\plotone{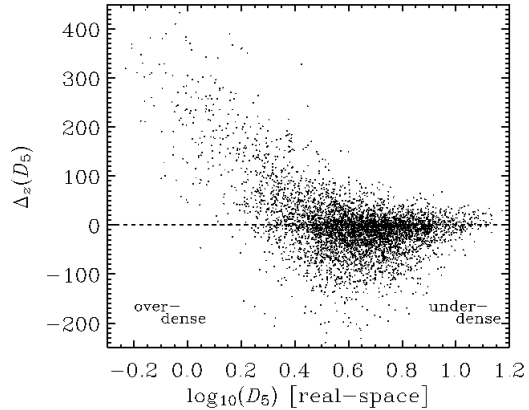}
\caption{We plot $\Delta_{z}(D_{5})$ (see eq.\ \ref{zseqn}) as a
function of $D_{5}$ as measured according to the real-space positions
of galaxies. The sample plotted includes the 5,031 galaxies in the
DEEP2-selected mock catalog with $0.7 < z < 1.4$ and located more than
$4\ h^{-1}$ comoving Mpc from a survey edge. Not surprisingly,
redshift-space distortions are most problematic in over-dense
environments. }
\label{zspace2}
\end{center}
\end{figure}

As illustrated in Figure \ref{zspace1}, the Voronoi volume and the
3-dimensional $3^{\rm rd}$- or $5^{\rm th}$-nearest-neighbor distances
are similarly affected by redshift-space distortions. For each
measure, the effects of redshift-space distortions are non-negligible
and as shown in Figure \ref{zspace2} are greatest in over-dense
environments. In comparison to the $V$ and 3-d $D_n$ measures, the
counts in an aperture density estimator, $C$, and projected $n^{\rm
th}$-nearest-neighbor measure, $D_{p,n}$, are less affected by the
``fingers-of-god'' due to their effective smoothing in the redshift
direction; by definition, the projected estimators, $C$ and $D_{p,n}$
forfeit sensitivity in the redshift direction, which reduces their
susceptability to redshift-space distortions. For nearly 80\% of the
sample, $C$ is unaffected $(\Delta C = 0)$ by peculiar motions when
using a cylindrical aperture with a length of $\pm 1000\ {\rm km}/{\rm
s}$ and diameter of $1\ h^{-1}$ comoving Mpc. The sensitivitity of $C$
to redshift-space distortions is somewhat dependent upon the choice of
the aperture size in the line-of-sight direction such that a smaller
aperture is more adversely affected. For more than 90\% of galaxies in
our sample, we find $|\Delta C| \le 1$, again using an aperture with
length of $\pm 1000\ {\rm km}/{\rm s}$. Similar results are found for
the projected $n^{\rm th}$-nearest-neighbor distance measure; more
than 80\% of the sample meets the criterion $\Delta_{z}(D_{p,n}) <
5\%$ for $n=3,5$ and using a velocity interval of $\pm 1000\ {\rm
km}/{\rm s}$ to exclude foreground and background sources.

\section{Target Selection and Observation}

There are inevitable trade-offs between the number density of sources
targeted for observation and the area of sky covered in a redshift
survey. Clustering of high-redshift galaxies and fiber or slit
collisions on multi-object spectrographs conspire to limit the
fraction of target objects which a survey can observe at one
time. Furthermore, not every object targeted will successfully yield a
redshift, generally due to finite signal-to-noise and instrumental
effects.

\begin{figure}[h]
\begin{center}
\plotone{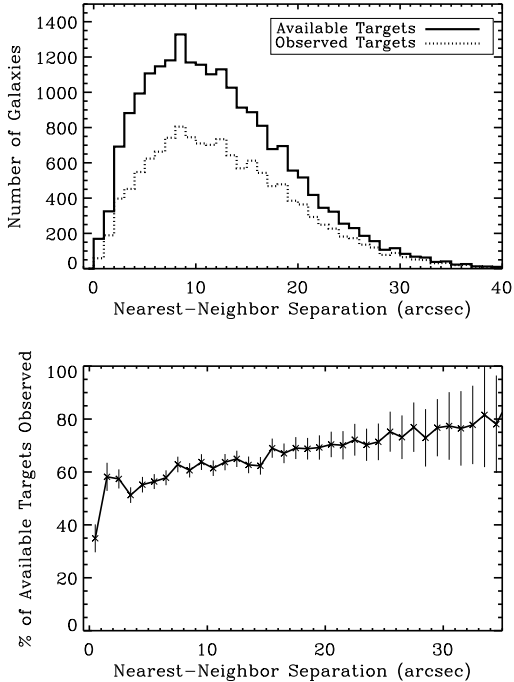}
\caption{(\emph{Top}) We plot the distribution of
angular-nearest-neighbor distances for galaxies in a $120' \times
30'$, DEEP2-selected mock catalog. The solid line gives the
distribution for all galaxies meeting the DEEP2 target-selection
criteria. The dotted line plots the distribution for galaxies in the
mock catalog which are selected for DEEP2 observation. (\emph{Bottom})
The percentage of available targets observed by DEEP2 as a function of
angular-nearest-neighbor distance. The error bars are given according
to Poisson statistics. The DEEP2 survey slightly under-samples dense
regions of sky.}
\label{target1}
\end{center}
\end{figure}

The DEEP2 redshift survey will target $\sim \! 50,000$ galaxies
covering $3.5$ square degrees of sky, over 80 nights on the Keck II
telescope \citep{davis04, faber05}. This impressive survey, however,
will only target approximately 60\% of available high-redshift $(0.7 <
z < 1.4)$ galaxies in its four fields and successfully measure
redshifts for about 75\% of targeted galaxies. The DEEP2 survey is
designed with the goals of studying large-scale structure and galaxy
properties at high redshift. Thus, the survey design attempts to
maximize the number of redshifts obtained, to sample the galaxy
distribution over a broad range of length scales, and to minimize the
effects of cosmic variance. Due to slit collisions on DEIMOS
slitmasks, the DEEP2 survey systematically under-samples regions of
sky with a high surface density of galaxies (see Figure
\ref{target1}). It is critical for a study of galaxy environments to
understand how this bias may affect the environment measured used.

While the detailed effects of target selection and redshift
incompleteness are clearly specific to each survey, the goal of this
section is to understand how a given environment measure is affected
by the limited sampling common to all deep redshift surveys. In this
work, we adopt the DEEP2 survey as a representative high-redshift,
spectroscopic survey. As discussed in \S 2, the DEEP2 survey targets
all galaxies at $R_{\rm AB} \le 24.1$ according to a probabilistic
algorithm which preferentially selects high-$z$ galaxies. Applying the
DEEP2 target-selection and slitmask-making algorithms to a
magnitude-limited $(R_{\rm AB} \le 24.1)$ mock catalog covering $40'
\times 30'$, the simulated survey targets and successfully measures
redshifts for 2,839 of the 5,866 galaxies in the field and with $0.7 <
z < 1.4$ (assuming a redshift success rate of $\sim \! 70\%$). To
study the combined effects of target selection, slitmask making, and
redshift success, we compute each environment measure on the
DEEP2-selected sample and on the full magnitude-limited sample.

\begin{figure}[h]
\begin{center}
\plotone{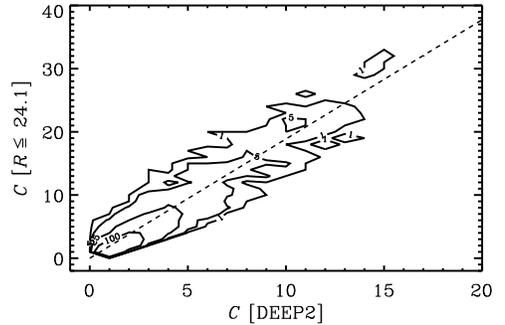}
\caption{We plot the relationship between counts in an aperture, $C$,
for the simulated DEEP2-selected redshift sample as measured on the
full magnitude-limited $(R_{\rm AB} \le 24.1)$ mock sample and as
measured on the subset of the magnitude-limited sample, successfully
observed by DEEP2. The contour levels plotted correspond to 1, 5, 25,
and 100 galaxies. The dashed line follows the redshift rate of the
DEEP2-selected mock sample $(\sim \! 50\%)$. We employ a cylindrical
aperture $2r_{\rm t}=1\ h^{-1}$ comoving Mpc in diameter and $\pm1000\
{\rm km}/{\rm s}$ along the line-of-sight. All galaxies with $D_{\rm
edge} < r_{\rm t}$ are excluded from the anaylsis so as to eliminate
edge effects. No evidence is found for any environment-dependent bias
in the DEEP2 target-selection and slitmask-making procedures.}
\label{cnts_fig}
\end{center}
\end{figure}

As illustrated in Figure \ref{cnts_fig}, the counts in an aperture
measure, $C$, shows no indication of an environment-dependent bias in
the DEEP2 target selection. If DEEP2 severely under-samples dense
environments, then we would expect to see a saturation in the observed
value of $C$ at high densities relative to the estimation of $C$
computed on the magnitude-limited sample. Instead, we find a linear
relation extending to dense environments which follows the $\sim \!
50\%$ overall completeness of the DEEP2-selected mock catalog.

\begin{figure}[h]
\begin{center}
\plotone{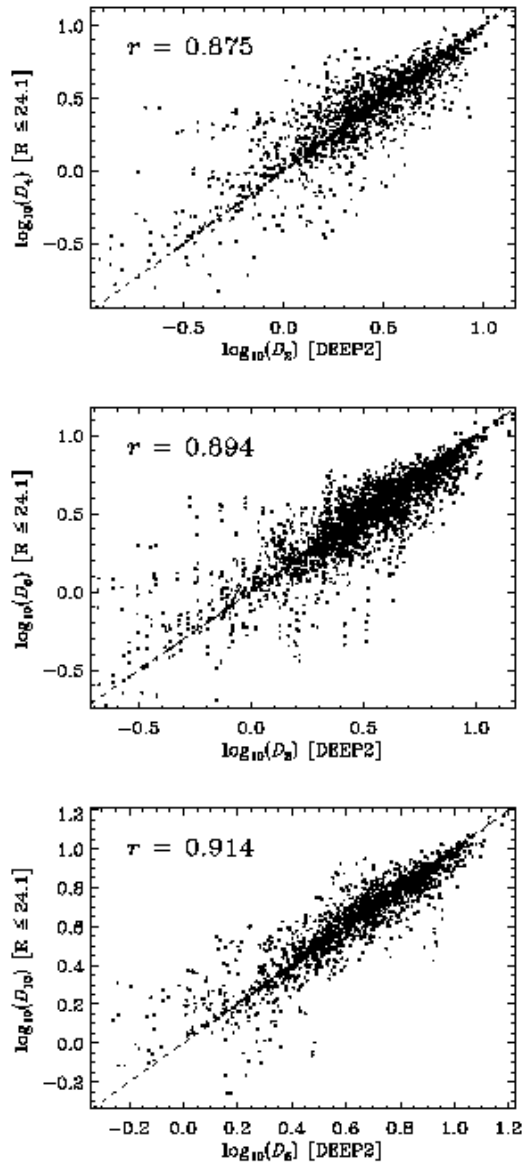}
\caption{(\emph{Top}) We plot the relation between $D_{2}$ as measured
on the observed DEEP2-selected galaxy sample and $D_{4}$ as measured
on the corresponding magnitude-limited $(R_{\rm AB} \le 24.1)$
sample. (\emph{Middle}) Plotted is the correlation between $D_{3}$ as
measured on the observed DEEP2-selected galaxy sample and $D_{6}$ as
computed on the corresponding magnitude-limited $(R_{\rm AB} \le
24.1)$ sample. (\emph{Bottom}) We plot $D_{5}$ computed from the
distribution of observed DEEP2-selected galaxies versus the $D_{10}$
values measured on the magnitude-limited data set. In each plot, the
Pearson correlation coefficient, $r$, is given in the upper left
corner and the dashed line follows a correlation of $r = 1$. Again, no
evidence is found for an environment-dependent bias in the DEEP2
target-selection and slitmask-making procedure.}
\label{mask1}
\end{center}
\end{figure}

\begin{figure}[h]
\begin{center}
\plotone{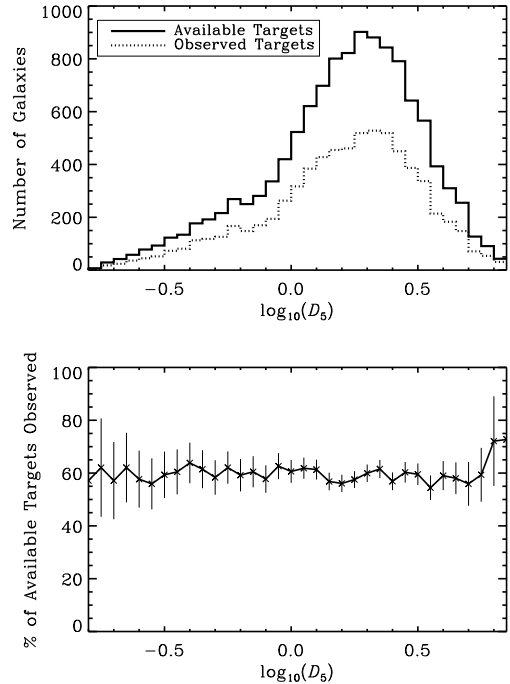}
\caption{(\emph{Top}) We plot the distribution of ``true'' environment
for DEEP2-selected galaxies in a $120' \times 30'$ mock catalog where
environment is traced by the 3-d $5^{\rm th}$-nearest-neighbor
distance as computed within a volume-limited sample using real-space
galaxy positons. The solid line gives the distribution for all
galaxies meeting the DEEP2 target-selection criteria. The dotted line
plots the distribution for galaxies in the mock catalog which are
selected for DEEP2 observation. (\emph{Bottom}) The percentage of
available targets observed by DEEP2 as a function of environment. The
error bars are given according to Poisson statistics. The DEEP2 survey
does not under-sample dense environments.}
\label{target3}
\end{center}
\end{figure}

Due to the fixed comoving aperture size of the counts in an aperture
environment measure, it probes the same physical scale independent of
how the targeted galaxies are selected. The $n^{\rm
th}$-nearest-neighbor distance measure, on the other hand, can sample
systematically different effective scales depending on the galaxy
sampling. As illustrated in Figure \ref{mask1}, within the observed
spectroscopic sample, the $5^{\rm th}$-nearest-neighbor distance is
roughly tracing the $10^{\rm th}$-nearest-neighbor distance in the
magnitude-limited mock; this is sensible, as the DEEP2-selected mock
samples $\sim 50\%$ of galaxies and what was the $10^{\rm th}$-nearest
neighbor in the magnitude-limited sample will typically be the $5^{\rm
th}$ in the DEEP2-selected sample. Similarly, the $2^{\rm nd}$- and
$3^{\rm rd}$-nearest-neighbor distances are effectively tracing the
$4^{\rm th}$- and $6^{\rm th}$-nearest-neighbor distances,
respectively, in the magnitude-limited mock (see Figure
\ref{mask1}). While target selection and slitmask making affect the
scale on which the $n^{\rm th}$-nearest-neighbor distance samples the
galaxy distribution, they do so in a manner which does not depend
upon environment. Thus, while DEEP2 under-samples dense \emph{regions
of sky}, the survey does not under-sample dense \emph{environments}
(see Figure \ref{target3}). 

The limited galaxy sampling of the DEEP2 survey causes the calculated
Voronoi volumes to be systematically larger than if computed on the
full magnitude-limited sample. We find that the level to which a given
Voronoi volume is affected by the DEEP2 sample selection does not
depend on $V$ or redshift; the limited sampling of the DEEP2 survey
simply introduces a random scatter towards larger volumes. Similar to
$D_n$ and $C$, we see no evidence for an environment-dependent bias
due to the DEEP2 target-selection procedures.


\section{Correcting for the Survey Selection Function}

For any magnitude-limited survey, the fraction of total galaxies in a
volume-limited sample within the magnitude limits varies -- commonly
decreasing -- as a function of redshift. This variable sampling of the
galaxy distribution as a function of redshift causes measurements of
galaxy densities to depend strongly on $z$. For instance, if a survey
under-samples at higher redshift, then estimates of $D_{n}$ and $V$ at
high $z$ will be artificially inflated relative to estimates at low
$z$; similarly $C$ will be under-estimated at higher redshift.  Often,
magnitude-limited redshift surveys are trimmed to a volume-limited
sub-sample to avoid these issues. At high redshift, however, this can
dramatically reduce the sample size; for example, selecting a
vol-limited sub-sample within a DEEP2-selected mock catalog excludes
as much as 40\% of the observed galaxies. Furthermore, over regimes
where luminosity evolution is significant $(\Delta_z \gtrsim 0.1)$,
even defining a volume-limited sample can be problematic.

To utilize the entire survey sample or for surveys that do not follow
a simple magnitude-limited target selection, the variations in the
galaxy sampling rate with redshift may be quantified in terms of a
survey selection function, $s(z)$, with which density estimates
(number of galaxies per comoving volume or number of galaxies per
projected comoving area) may be corrected as follows:
\begin{equation}
{ X_{0}(\alpha, \delta, z) = \frac{ X_{z}(\alpha, \delta, z) }
{  s(z) \cdot w(\alpha, \delta) } }\ ,
\label{sofz_eq}
\end{equation}
where $X_{z}$ is the density estimate computed from the observed
redshift distribution, $w$ is a 2-dimensional survey completeness map
which accounts for variation in redshift completeness from field to
field within the survey, and $X_0$ is the corrected density estimate.

There are several ways to determine the selection function of a
survey. The most common approach is to first estimate the galaxy
luminosity function (LF) for all of the galaxies in the redshift
survey, and then use it to predict the redshift distribution of the
sample \citep[e.g.][]{madgwick03}. However, unless evolution is
correctly incorporated, the LF will not alone be able to correctly
predict the redshift dependence of the observed number density of
galaxies in a deep survey. Furthermore, working at high redshift, it
becomes increasingly difficult to constrain the galaxy LF since
observations are limited to the brightest sources, thereby making
estimations of the characteristic luminosity, $M_{*}(z)$, and
faint-end slope, $\alpha(z)$, less secure \citep[e.g.][]{willmer05,
wolf03, bell04}. For this reason, the selection function for a survey
is often estimated by smoothing the observed number density of
galaxies, $n(z)$, as a function of redshift and then normalizing
according to an assumed dependence of the comoving number density of
galaxies on $z$ \citep[e.g.][]{coil04a}. This has the disadvantage
that density inhomogeneities in the survey will somewhat affect the
derived redshift distribution, even with large smoothing kernels, due
to the strength of cosmic variance; also, kernels large enough to
minimize this will distort real features in $n(z)$, especially where
there are large gradients. On the other hand, any evolution in the
observed number density of galaxies with $z$ will be automatically
incorporated into the estimation of $s(z)$.

Yet another approach to estimating $s(z)$ is to compute an analytical
fit to the observed data from which the selection function is then
derived \citep[e.g.][]{cooper05b, faber05}. Similar to a selection
function estimated from smoothing the observed $n(z)$ distribution, an
analytical fit to the data -- or ``fitting'' method for estimating
$s(z)$ -- is subject to cosmic variance, but to a much smaller degree
than the ``smoothing'' method, as small-scale variations in $n(z)$
which do not match the model are not allowed.

In this work, we estimate the survey selection function for the mock
DEEP2 survey according to four different prescriptions: (a) estimating
$s(z)$ by smoothing the observed $n(z)$ distribution in a
DEEP2-selected mock catalog $(120' \times 30')$ assuming no evolution
in the comoving number density of galaxies with redshift, using a
similar algorithm as \citet{coil04a} (``smoothing'' method), (b)
fitting for the selection function assuming a functional form for the
redshift dependence of the successfully-observed $dN/dz$ and again
assuming no evolution in the comoving number density of galaxies
(``fitting'' method), (c) determining the true selection function by
computing the number density of available targets over many DEEP2 mock
pointings relative to the volume-limited number density of galaxies in
the mocks, and (d) deriving $s(z)$ from the conditional LF assumed in
constructing the mock catalogs. This last approach is identical to the
commonly-used method of estimating the selection function using the
measured LF and predicting the redshift distribution of the underlying
galaxy population. The first two methods, (a) and (b), are analagous
to methods one might use to derive $s(z)$ solely from the
observational data in a deep redshift survey and are subject to cosmic
variance and uncertainties in the assumed normalization and redshift
dependence of the comoving number density of galaxies. The latter two
methods are effectively identical, and test that the mock catalogs are
working as advertised.

\begin{figure}[h]
\begin{center}
\plotone{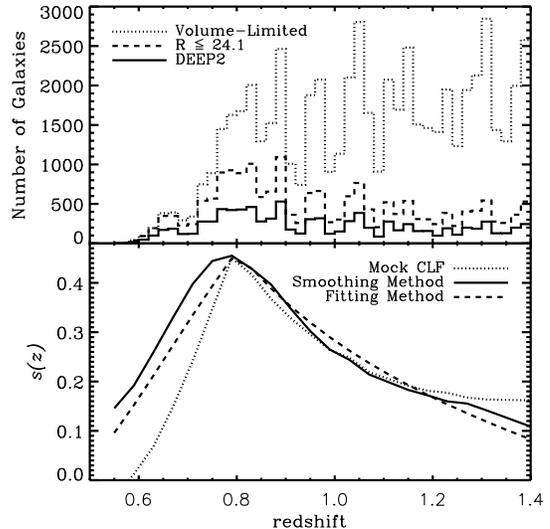}
\caption{ (\emph{Top}) We plot the galaxy redshift distributions for
the mock catalogs. The solid line plots the observed redshift
distribution, $N(z)$, for galaxies within a DEEP2-selected mock
catalog $(120' \times 30')$. The dotted line gives the volume-limited
distribution of redshifts for all galaxies $(L > 0.1 L_{*})$ in the
same $120' \times 30'$ field. A probablistic roll-off in number
density has been introduced into the mocks at $z < 0.8$ to simulate
the high-$z$ color-selection used in the DEEP2 target-selection. The
dashed line shows the magnitude-limited $(R_{\rm AB} \le 24.1)$
redshift distribution drawn from the same 1 square degree
field. (\emph{Bottom}) The solid line gives the selection function,
$s(z)$, derived from successively smoothing the observed $n(z)$
distribution in the DEEP2-selected mock DEEP2 catalog on scales of
$\Delta z = 0.15$ and assuming that $\sim \!\!  45\%$ of all available
targets are successfully observed. 
The dashed line shows the analytical fit for the selection function as
derived from the same DEEP2-selected mock. Lastly, the dotted line
gives the estimation of $s(z)$ derived from the conditional LF
employed in constructing the mock catalogs. Note that all of the
$s(z)$ curves have been normalized to peak at $s(z)=0.45$.}
\label{sofz_fg}
\end{center}
\end{figure}

In Figure \ref{sofz_fg}, we present the mock selection functions
derived using each of the methods described above. In general, the
agreement between the different approaches for determining $s(z)$ is
quite good. At the highest redshifts $(z > 1.0)$, the different
estimations of the selection function differ due to differences in the
assumed comoving number density of galaxies at high redshift. The
conditional LF adopted in constructing the mock catalogs yields
evolution that produces a decrease in the comoving number density,
$\nu(z)$, of galaxies at $z > 1.0$, while in estimating $s(z)$ from
the observed mock DEEP2-selected redshift distribution we assume a
constant form for $\nu(z)$. Both estimations of the comoving number
density at $z > 1.0$ are consistent with existing observational
evidence \citep{wolf03, willmer05}.

The footprint of large-scale structure on a selection function derived
from smoothing the observed DEEP2-selected mock $n(z)$ distribution is
reduced if a large smoothing kernel is used; here, we apply two
successive smoothing windows of width $\Delta z = 0.15$. If the
smoothing kernel is too small, the presence or absence of structures
such as filaments or walls (i.e. cosmic variance) will cause us to
overestimate or underestimate the fraction of galaxies sampled at a
given redshift. Accordingly, over- or under-densities of galaxies will
be inappropriately reduced in amplitude when corrected by the survey
selection function; e.g. the presence of a filament will push the
measured $s(z)$ up at its redshift, reducing the corrected density
measured, $X_0$, artificially. Any smoothing large enough to erase the
effects of cosmic variance in a survey covering a few square degrees
will, unfortunately, cause flattening in the shape of $s(z)$,
especially near the limits of the redshift range probed. Due to the
drawbacks of smoothing, fitting for the selection function as detailed
above is often a superior method for estimating $s(z)$ from the
observed data, but it does require assumptions about the form of
$dN/dz$, which smoothing does not.

To study the effectiveness of correcting the measured galaxy densities
by the factor of $1 / s(z)$ (see eq.\ \ref{sofz_eq}), we have
computed the projected $7^{\rm th}$-nearest-neighbor surface density,
$\Sigma_7$, within a volume-limited mock catalog covering $120' \times
30'$ of sky. We then compare this to the projected $3^{\rm
rd}$-nearest-neighbor surface density, $\Sigma_3$, for those galaxies
successfully observed in the DEEP2-selected sample. We then correct
these ``observed'' $\Sigma_3$ values using each of the $s(z)$ shown in
Figure \ref{sofz_fg}, and also attempt an empirical correction. This
correction is given by dividing each observed density value by the
median $\Sigma_3$ for galaxies at that redshift where the median is
computed in a bin of $\Delta z = 0.04$. Correcting the measured
densities in this manner converts the $\Sigma_3$ values into measures
of over-density relative to the median density and is similar to the
methods employed by \citet{hogg03} and \citet{blanton03b}.

\begin{figure}[h]
\begin{center}
\plotone{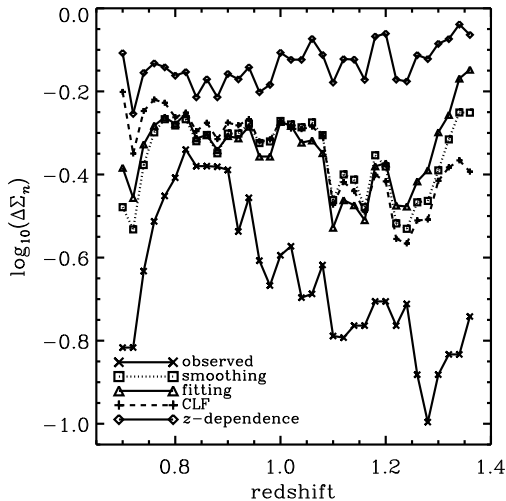} 
\caption{We plot the difference in the median density measured in a
DEEP2-selected mock catalog -- using the various $s(z)$ estimations to
correct the observed $\Sigma_3$ values -- relative to the median
density, $\Sigma_7$, computed in the corresponding volume-limited
sample, as a function of $z$. Ideally, these curves should be
flat. The median densities are computed in redshift bins of width
$\Delta z = 0.02$. At redshifts $z < 0.8$, the $\Sigma_7$ values
measured in the volume-limited sample decrease due to the fall-off in
$\nu(z)$ forced in the mock catalogs. For this reason, comparisons of
the various $s(z)$ are not valid at low redshift. The empirical
correction (labeled as {\it z-dependence}) is at least as effective as
correcting using a selection function.}
\label{sofz_comp} 
\end{center} 
\end{figure}

Figure \ref{sofz_comp} illustrates the effectiveness of each selection
function at reproducing the redshift dependence of the galaxy density
distribution as measured in the volume-limited sample. Within redshift
bins of $\Delta z = 0.02$, we compute the difference in median density
between the corrected $\Sigma_3$ values and the median density,
$\Sigma_7$, of objects in the volume-limited sample. While each of the
methods for estimating the survey selection function is an improvement
over the uncorrected density distribution, an empirical correction (as
described in the previous paragraph) which removes all $z$-dependence
in the observed density distribution is at least as effective as
correcting using a selection function (see Table \ref{sofz_tab}).

\begin{deluxetable}{c c c}
\tablewidth{0pt}
\tablecolumns{3}
\tablecaption{\label{sofz_tab} Effectiveness of selection functions}
\tablehead{ $s(z)$    & rms        &    $\rho$  }
\startdata
observed  & 0.180      &  0.576      \\
smoothing & 0.094      &   0.608          \\
fitting   & 0.083      &   0.603          \\
mock CLF  & 0.096      &   0.608         \\
z-dependence & 0.049   &   0.606 \\
\enddata
\tablecomments{For each determination of the selection function, we
compute the Spearman ranked correlation coefficient, $\rho$, between
the corrected $3^{\rm rd}$-neareast-neighbor surface density $\Sigma_3
/ s(z)$ measured in a DEEP2-selected mock catalog and the $7^{\rm
th}$-nearest-neighbor surface density measured in a volume-limited
sample. We also present the RMS fluctuations for each relation plotted
in Figure \ref{sofz_comp} over the redshift range $z > 0.8$. Each of
the selection functions improves the correlation $(\rho)$ between the
measured environment within the DEEP2-selected sample and the ``true''
environment measured in the volume-limited sample; the empirical
correction ({\it z-dependence}) has the smallest residual scatter with
redshift (partially by construction).}
\label{sofz_tab}
\end{deluxetable}


\section{Discussion}

Every environment measure that we have considered has its advantages
and disadvantages. The counts in an aperture measure, $C$, lacks
sensitivity in low-density environments and while not lacking in
dynamic range, it provides a non-continuous (or quantized) measure of
environment, a particular disadvantage if the typical value of $C$ is
small. It is best suited for working in dense environments where $C$
is more robust to redshift-space distortions than other measures and
for analyses in which one wishes to classify a sample into coarse
density bins or classes. The counts in an aperture statistic is unique
among the environment estimators tested in that it measures the galaxy
density on a clearly defined, fixed length scale. In contrast, the
projected and 3-dimensional $n^{\rm th}$-nearest-neighbor distance
measures probe the radius enclosing some total number of galaxies and
are not direct density measures. The $C$ parameter also provides a
great advantage via its robustness to survey edge effects.

\begin{figure}
\begin{center}
\plotone{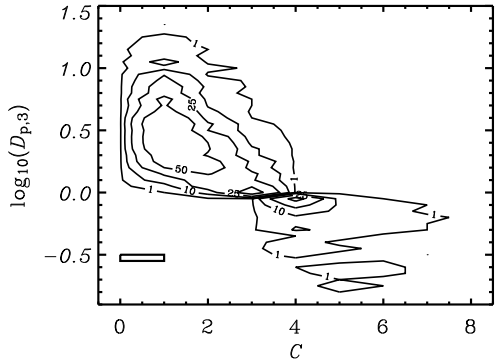}
\caption{We plot the correlation between projected $3^{\rm
rd}$-nearest-neighbor distance, $D_{p,3}$, and the counts in a
cylindrical aperture, $C$, measured in a DEEP2-selected mock catalog
$(40' \times 30')$. Here, we restrict the plot to those galaxies
removed from any survey edge by $(D_{\rm edge} > 2\ h^{-1}\ {\rm
comoving}\ {\rm Mpc})$. The contour levels plotted correspond to
$N=1,10,25,50$ and were computed using a sliding box as shown in the
lower-left corner of the plot. The counts in an aperture utilize a
cylindrical aperture of scale $2r_{\rm t}=1\ h^{-1}$ comoving Mpc in
diameter and $\pm1000\ {\rm km}/{\rm s}$ along the
line-of-sight. Similarly, a velocity interval of $\pm1000\ {\rm
km}/{\rm s}$ is used in restricting foreground and background galaxies
in the computation of $D_{p,3}$. In under-dense regions, the $C$
values are found to saturate, thereby limiting the utility of the
statistic.}
\label{DpvsC}
\end{center}
\end{figure}

Similar to the counts in an aperture statistic, the projected $n^{\rm
th}$-nearest neighbor distance measure is well suited for measuring
density in groups and clusters. However, unlike $C$, the projected
$D_{p,n}$ parameter provides a continuous estimate of galaxy density
extending to under-dense environments where it still provides a
reasonably accurate measure. While slightly more robust to edges than
its 3-dimensional counterpart, $D_n$, the projected $n^{\rm
th}$-nearest-neighbor distance is more prone to survey edge
contamination than the counts in an aperture statistic. Figure
\ref{DpvsC} shows the correlation between $D_{p,3}$ and $C$ as
computed in a $40' \times 30'$ simualted DEEP2 pointing. The
saturation of $C$ in less-dense regions is striking and proves to be a
significant drawback for a density estimator which is otherwise
extremely robust to survey edges and redshift-space distortions.

\begin{figure}[h]
\begin{center}
\plotone{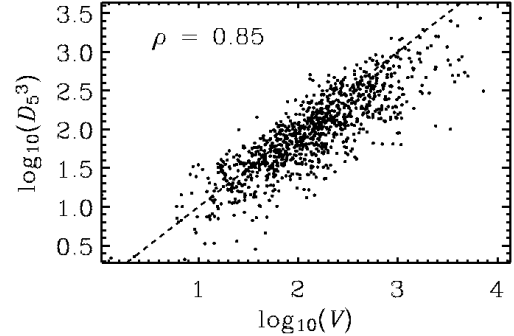}
\caption{We plot the correlation between Voronoi volume, $V$, and 3-d
$5^{\rm th}$-nearest-neighbor distance, $D_5$, measured on a
DEEP2-selected mock field $(120' \times 30')$. Here, we restrict the
plot to those galaxies far removed from any survey edges, $(D_{\rm
edge} > 5\ h^{-1}\ {\rm comoving}\ {\rm Mpc})$. Note that the $D_5$
values have been cubed to facilitate comparison to the Voronoi
volumes. The Spearman ranked correlation coefficient, $\rho=0.85$,
quantifies the strong agreement between the different density
estimators.}
\label{VvsD}
\end{center}
\end{figure}

The 3-dimensional $n^{\rm th}$-nearest neighbor distance and Voronoi
volume statistics are the best suited for measuring under-dense
environments. In groups and clusters, however, these density
estimators are significantly affected by redshift-space distortions,
causing each measure to become saturated. As illustarted in Figure
\ref{VvsD}, far removed from survey
edges, the Voronoi volume and $D_5$ measures agree very well over all
environments observed in the DEEP2-selected mock catalogs. However,
for the simulated DEEP2 survey data, the $n^{\rm th}$-nearest-neighbor
distance is much more robust to edge effects and is less expensive to
calculate.

For studies of environment at high redshift, including analysis in the
DEEP2 survey, we conclude that among the environment measures tested
the projected $n^{\rm th}$-nearest-neighbor distance provides the most
accurate estimate of local galaxy density over the broadest range of
scales. For work in dense environments, the $D_{p,n}$ offers great
robustness to redshift-space distortions and maintains a reasonably
high level of accuracy in under-dense environments. While $D_{p,n}$
can be affected by survey edges, contamination from geometric
distortions is easily understood and effectively minimized without
dramatically reducing the galaxy sample.


\section{Conclusions}

We have studied the applicability of several galaxy-density estimators
within deep redshift surveys at $z \sim 1$ utilizing the mock galaxy
catalogs of \citet{yan04}. We conclude as follows:

\begin{enumerate}

\item Photometric redshifts derived from multi-band photometry
$(\sigma_z \gtrsim 0.02)$ are not suitable for measuring galaxy
densities. Current photometric redshift surveys such as COMBO-17 do
not have the redshift precision needed to study environment at high
redshift. While more costly to obtain, spectroscopic redshifts are
requisite to accurately probe the local galaxy environment in a large
survey.

\item With the exception of the counts in an aperture estimator, $C$,
survey field edges are a major source of contamination for each
environment measure tested. To reduce these edge effects without
biasing the sample, all galaxies within some comoving distance ($\sim
\! 1-2\ h^{-1}$ comoving Mpc for DEEP2) of a transverse survey edge
should be rejected. At $z \sim 1$, excluding all galaxies within $1\
h^{-1}$ comoving Mpc of an edge removes roughly 0.05 degrees along
each dimension of the survey field. For smaller high-redshift surveys,
such as CFRS \citep{lilly95} or CNOC2 \citep{yee00}, edge effects
introduce contamination to a considerable portion of the survey data
set, thereby limiting the statistical power of the samples. Likewise,
for a survey of the GOODS-North field \citep{giavalisco04}, edge
effects would bias density measurements over $\sim \! 75\%$ of the
field. Testing each environment measure on a simulated DEEP2-selected
mock sample $(40' \times 30')$, the Voronoi volume is most severely
affected by edges, with more than 2 times as much contamination from
edge effects than $D_n$ or $D_{p,n}$. The counts in an aperture
measure displays the best behavior near edges of a survey field, with
a nearly negligible portion of the sample contaminated in our
simulations.

\item Redshift-space distortions are a significant and fundamental
roadblock to measuring accurate galaxy densities in over-dense
environments. The $n^{\rm th}$-nearest-neighbor distance measured in
3-dimensions and the Voronoi volume are most greatly affected, while
estimators such as projected $n^{\rm th}$-nearest-neighbor distance
and counts in an aperture -- which smooth the galaxy distribution
along the line-of-sight -- are less affected by the
``fingers-of-god''. Still, it should be noted that less than 15\% of a
simulated $R_{\rm AB} \le 24.1$ galaxy sample occupies environments at
which the $V$ and $D_n$ statistics saturate due to redshift-space
distortions.

\item The target selection algorithm employed by a survey 
could lead to environment-dependent biases in the observed galaxy
sample. The DEEP2 survey, which slightly under-samples dense regions
of sky, is equally sensitive at high and low densities. That is, we
find that the DEEP2 survey equally samples all environments at $z \sim
1$ (see Figure \ref{target3}). Also, we find that while the DEEP2
survey samples only $\sim 50\%$ of galaxies at $z \sim 1$, this
uniform incompleteness simply introduces a random scatter in the
measured environments and does not introduce an environment-dependence
bias.

\item In examining the evolution of galaxy environments as a function
of redshift, the estimation of the survey selection function plays a
critical role. Uncertainties in the comoving number density of
galaxies at high $z$ make comparisons over large redshift intervals
$(\Delta z \sim 0.5)$ problematic. Apart from such ambiguities, simple
empirical corrections for densities as a function of redshift are
highly effective.

\item For the DEEP2 Galaxy Redshift Survey, the projected $n^{\rm
th}$-nearest-neighbor distance provides the most accuracte estimate of
local galaxy density over a continuous and broad range of scales. The
$D_{p,n}$ statistic is reasonably robust to redshift space distortions
and still effective at tracing galaxy environments in under-dense
regions.

\item Among current data sets at high redshift, we find the DEEP2
Galaxy Redshift Survey provides the best opportunity for measuring
accurate galaxy environments over a broad and continuous range of
scales. The high sampling rate and excellent redshift-precision of
DEEP2 enable environments to be measured in even the most over-dense
regions and yield improved accuracy over other deep
surveys. Furthermore, DEEP2's high-precision redshifts and large
survey area (3.5 square degrees) minimize the effects of edges in both
the transverse and line-of-sight directions.

\end{enumerate}


\vspace*{0.05in}

\acknowledgements We wish to thank Chris Marinoni for providing his
Voronoi-Delaunay method group-finding code. This work was supported in
part by NSF grant AST00-71048. JAN and DSM acknowledge support by NASA
through Hubble Fellowship grants HST-HF-01165.01-A and
HST-HF-01163.01-A, respectively, awarded by the Space Telescope
Science Institute, which is operated by AURA Inc.\ under NASA contract
NAS 5-26555. BFG acknowledges support from a NSF Fellowship. MCC
thanks Mike Blanton for useful discussions about this work. MCC also
thanks Josh Simon and Alison Coil for careful reading of this
manuscript and many insightful suggestions which have improved this
work. Lastly, the authors are humbly indebted to Steve Dawson for his
invaluable assistance with various technical aspects of this effort.



\end{document}